\documentclass[11pt]{article}
\newcount\biblio
\title{Rigid motions:
action-angles, relative cohomology and polynomials
with roots on the unit circle}
\author{\small\textsc{J.-P. Fran\c{c}oise}\\
\small Universit\'e P.-M. Curie,\\
\small Laboratoire J.-L. Lions,\\ UMR 7598 CNRS\\
\small 4 Pl. Jussieu, 75252 Paris\\
\small France
\and \small\textsc{P.L. Garrido}
\\
\small Institute Carlos I for \\
\small Computational and \\
\small Theoretical Physics, \\
\small Universidad de Granada, \\
\small Espa$\tilde{\rm n}$a
\and\small\textsc{G. Gallavotti}
\\
\small Dipartimento di Fisica and INFN,\\
\small Universit{\`a} di Roma "La Sapienza",\\
\small Italia}%

\usepackage{ifthen}
\usepackage{hyperref}

\let\a=\alpha \let\b=\beta  \let\g=\gamma  \let\d=\delta \let\e=\varepsilon
\let\z=\zeta  \let\h=\eta   \let\th=\theta  \let\l=\lambda
\let\m=\mu        \let\x=\xi     \let\p=\pi    
\let\s=\sigma    \let\f=\varphi 
  \let\ps=\psi  \let\o=\omega
\let\G=\Gamma  \let\Th=\Theta \let\X=\Xi

\let\wt=\widetilde
\def\wh{\widehat}
\let\0=\noindent
\def\defi{\,{\buildrel def\over=}\,}

\def\UU{{\cal U}}
\def\HH{{\cal H}}
\def\otto{\,{\kern-1.truept\leftarrow\kern-5.truept\to\kern-1.truept}\,}

\def\KJ{{\bf K}}\def\dn{{\,{\rm dn}\,}}\def\sn{{\,{\rm sn}\,}}
\def\cn{{\,{\rm cn}\,}}
\def\*{\vskip2mm}
\def\lis#1{{\overline#1}}
\def\Eq#1{{\label{#1}}}%
\def\equ#1{(\ref{#1})}
\def\be{\begin{equation}}\def\ee{\end{equation}}
\def\tende#1{\,\vtop{\ialign{##\crcr\rightarrowfill\crcr
 \noalign{\kern-1pt\nointerlineskip} \hskip3.pt${\scriptstyle
 #1}$\hskip3.pt\crcr}}\,}
\def\V#1{{\bf#1}}

\def\iniz{\setcounter{equation}{0}}
\renewcommand{\theequation}{\arabic{section}.\arabic{equation}}

\def\inizA{\setcounter{equation}{0}{%
\renewcommand{\theequation}{\Alph{section}.\arabic{equation}}}}

\def\eqalign#1{\null\,\vcenter{\openup\jot
  \ialign{\strut\hfil$\displaystyle{##}$&$\displaystyle{{}##}$\hfil
      \crcr#1\crcr}}\,}
\def\bea{\begin{eqnarray}}
\def\eea{\end{eqnarray}}
\def\nn{\nonumber}
\newdimen\xshift \newdimen\xwidth \newdimen\yshift \newdimen\ywidth

\def\ins#1#2#3{\vbox to0pt{\kern-#2pt\hbox{\kern#1pt #3}\vss}\nointerlineskip}

\def\eqfig#1#2#3#4#5{
\par\xwidth=#1pt \xshift=\hsize \advance\xshift
by-\xwidth \divide\xshift by 2
\yshift=#2pt \divide\yshift by 2
{\hglue\xshift \vbox to #2pt{\vfil
#3 \includegraphics{#4.eps}
}\hfill\raise\yshift\hbox{#5}}}

\def\8{\write12}  %abbreviazione{\openout15=\jobname.aux}

\begin{document}
\maketitle

\begin{abstract}\0Revisiting canonical integration of the classical
    solid near a uniform rotation, canonical action angle coordinates,
    hyperbolic and elliptic, are constructed in terms of various power
    series with coefficients which are polynomials in a variable $r^2$
    depending on the inertia moments. Normal forms are derived via the
    analysis of a relative cohomology problem and shown to be obtainable
    without the use of ellitptic integrals (unlike the derivation of the
    action-angles). Results and conjectures also emerge about the
    properties of the above polynomials and the location of their roots.
    In particular a class of polynomials with all roots on the unit circle
    arises.\end{abstract} \*

\0{\small {\bf Key words:} \it Elliptic Functions, Gyroscope, Solid,
    Deprit angles, Canonical Integrablity, Relative Cohomology, Polynomials
Roots\ }%
\footnote[1]{\small\texttt{jpf@math.jussieu.fr},
\texttt{garrido@onsager.ugr.es},\\
\hglue.6cm\texttt{giovanni.gallavotti@roma1.infn.it}
}%
\*

\section{Overview}\label{sec1}
\iniz Integration of a rigid body motions is well known.  In this paper we
recall in Sec.\ref{sec2} the kinematic description of the coordinates in
which the system is obviously integrable by reduction to quadratures
(Deprit's coordinates, described in Eq.\equ{e2.5}). We try to treat
simultaneously the motions near all proper rotations, at the expense of
having to append labels $a=\pm1$ to most quantities thus making the
notation a little heavier. This is done to keep always clear that the
stable and unstable motions are in some sense described by the same
analytic functions.

Our first aim is to obtain canonical coordinates $p,q$, in the vicinity of
the proper rotations, such that the Hamiltonian, at fixed total angular
momentum $A$, becomes a function of the product $pq$ (near unstable proper
rotations) or of $p^2+q^2$ (near the stable ones).  Although existence of
such coordinates is well known and their determination is in principle
controlled by appropriate (elliptic) integrals, \cite[Sec.50]{Ar989}, it
remains, to our knowledge, only implicit in the literature,
\cite[Sec.69]{Wh917}.

In Sec.\ref{sec3} explicit expressions for the elliptic integrals are
derived, Eq.\equ{e3.7}, which give the motions in terms of variables
$p',q'$ evolving exponentially in time on the hyperbolae $p'q'=const$ or
rotating uniformly on the circles $p'^2+q'^2=const$ and ``reducing'' the
problem to determine the nontrivial scaling function $C$ (depending on
$p'q'$ or $p'^2+q'^2$) which leads to the above mentioned canonical
coordinates via $p=p'C$, $q=q'C$. In Appendix \ref{AppA} given details on
the computation of appropriate elliptic integrals: this should also clarify
the advantages of the cohomological analysis of the following
Sec.\ref{sec4} and \ref{sec6}.

In Sec.\ref{sec4} the scaling function $C$ is related to a Jacobian
determinant between two differential forms, and it is computed: this is
done, avoiding evaluation of other elliptic integrals, by computing the
cohomology of the forms $dB\wedge d\b$ (where $B,\b$ are two Deprit's
coordinates, see Sec.\ref{sec2}) and $dp'\wedge d q'$ relative to the
functions $p'q'$ or $p'^2+q'^2$. The canonical coordinates are therefore
completely determined, Eq.\equ{e4.10},\equ{e4.11} (together with
Eq.\equ{e3.7}). 

The latter expressions are still quite implicit and in Sec.\ref{sec5} we
bring them to a form suitable for the evaluation of the integrating
coordinates via computable power series; and as an example a few terms of
the scaling functions $C$ and of the normal form of the Hamiltonian are
evaluated, Eq.\equ{e5.4}, \equ{e5.6}. Doing so various families of
polynomials, in a variable $r^2$ depending on the inertia moments, arise.
The polynomials appear to have zeros on very special locations suggesting
possible conjectures. 

To understand the properties of the polynomials which arise in the
construction of the normal forms we show, Sec.\ref{sec6}, that the normal
form can be generated by solving another problem of relative cohomology:
this leads to an explicit determination of the normal forms in terms of a
further family of polynomials. In this case we prove that some of the new
polynomials have all roots on the unit circle (with the help of theorems on
the location of zeros of symmetric polynomials): and we conjecture a
relation with the Lee-Yang theorem for the zeros of ``ferromagnetic
polynomials''.

In Sec.\ref{sec7} we determine the action angle coordinates for the other
two Deprit variables (which either do not appear in the
Hamiltonian or are constants of motion).

In Sec.\ref{sec8} we apply the relative cohomology method of
Sec.\ref{sec4} and \ref{sec6} to the pendulum, to illustrate the
cohomological method. And to the geodesic flow on revolution ellipsoids,
with the purpose of showing that in the latter case also appear natural
families of polynomials in a parameter $r^2$ (depending on the ellipsoid
equatorial and polar axes): but in this case the zeros are not located on
the unit circle, {\it although almost so}: Eq.\equ{e8.6}.

Appendices contain supplements ({\it e.g.} the proof of Chen's
theorem).

%%%%%%%%%%%%%%%%%%%%%%%%%%%%%%%%%%%%%%%%
%%%%%%%%%%%%%%%%%%%%%%%%%%%%%%%%%%%%%%%%
\section{Solid with fixed center of mass}\label{sec2}
\iniz
%%%%%%%%%%%%%%%%%%%%%%%%%%%%%%%%%%%%%%%%
%%%%%%%%%%%%%%%%%%%%%%%%%%%%%%%%%%%%%%%%

The theory of Jacobian elliptic functions, for reference see \cite{GR965},
yields a complete calculation for the motion of a solid with a fixed point.
This is revisited here, to exhibit a few interesting properties of the
relevant elliptic integrals.

The Hamiltonian of a solid with inertia moments $I_1,I_2,I_3$, in Deprit's
canonical coordinates $(K_z, A, B,\g,\f,\b)$, see problems in Sec.4.11 of
\cite{Ga983}, is
\be\wt H(K_z,A,B,\g,\f,\b)=\frac12
\frac{B^2}{I_3}+\frac12\big(\frac{\sin^2\b}{I_1}+
\frac{\cos^2\b}{I_2}\big)\,(A^2-B^2).\Eq{e2.1}\ee
and it will be convenient to order the proper axes of the solid body so
that $I_3<I_1<I_2$.  In this way the motions with $B=0$ and $\b=0,\p$ or
$\pm\frac\p2$ are stable or unstable rotations, respectively.  For
convenience the unstable rotations will be studied by shifting by
$\frac\p2$ the origin of the angle $\b$, so that for the unstable rotations
we shall use the Hamiltonian:
\be\wt H(K_z,A,B,\g,\f,\b)=\frac12
\frac{B^2}{I_3}+\frac12\big(\frac{\cos^2\b}{I_1}+
\frac{\sin^2\b}{I_2}\big)\,(A^2-B^2).\Eq{e2.2}\ee
Motions can be described by referring them to the two node lines $\V m$ and
$\V n$: $\V m$ being the node between the angular momentum plane,
orthogonal to the angular momentum $\V M=A\, \V k$, and the fixed reference
plane $\lis {{\bf i}}-\lis {{\bf j}}$, and $\V n$ being the node between
the angular momentum plane and the inertial plane $1-2$.  The angle $\g$
locates the node $\V m$ on the fixed plane and the angle $\f$ locates $\V
n$ on the angular momentum plane with respect to $\V m$.

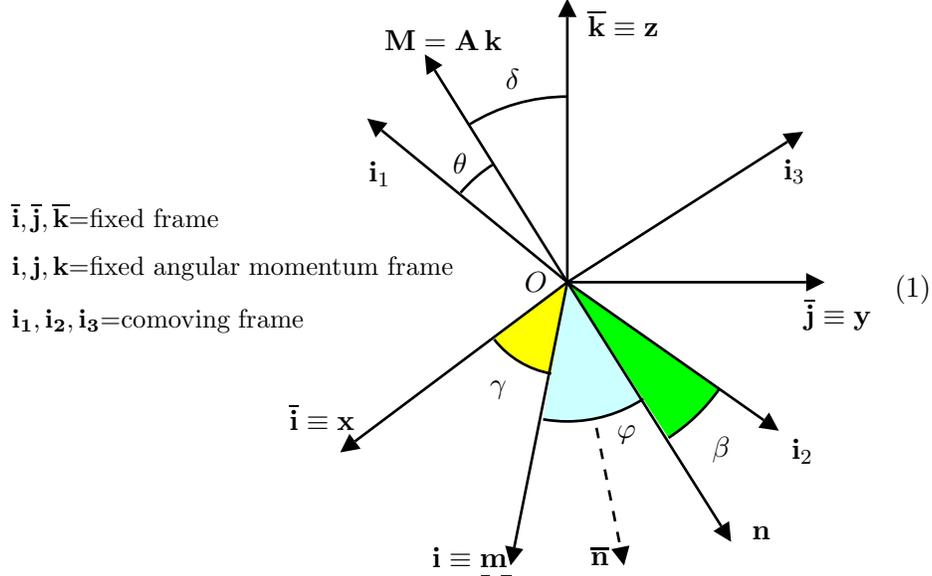
\vglue6mm
%%%\eqfig{190}{190}
\eqfig{90}{190}
{
\ins{-140}{130}{\small ${\bf \lis i,\lis j, \lis k}$=fixed frame}
\ins{-140}{110}{\small ${\bf i, j,  k}$=fixed angular momentum frame}
\ins{-140}{90}{\small ${\bf i_1,i_2,i_3}$=comoving frame}
\ins{-35 }{54 }{${\bf \lis i\equiv x}$}
%\ins{22 }{34 }{${\bf \lis \f}$}
\ins{41 }{63 }{${\bf \g}$}
\ins{19 }{0 }{${\bf i}\equiv{\bf m}$}
\ins{79 }{0 }{${\bf \lis n}$}
\ins{89 }{46 }{${\f}$}
%%%\ins{98  }{72 }{${\psi}$}
\ins{125}{42 }{${\b}$}
\ins{140 }{8 }{${\bf n}$}
%%%\ins{155 }{41 }{${\bf i}_1$}
\ins{155 }{41 }{${\bf i}_2$}
%\ins{130 }{37 }{${\lis\psi}$}
%%%\ins{152 }{147 }{${\bf i}_2$}
\ins{152 }{147 }{${\bf i}_3$}
\ins{78 }{203 }{${\bf\lis k\equiv z}$}
\ins{1 }{196 }{${\bf M=A\,\V k}$}
%%%\ins{-5 }{146 }{${\bf i}_3$}
\ins{-5 }{146 }{${\bf i}_1$}
\ins{47 }{181 }{${\d}$}
\ins{27 }{149 }{${\th}$}
%\ins{54 }{146 }{${\lis\th}$}
\ins{54 }{104 }{$O$}
\ins{160 }{94 }{${\bf \lis j\equiv y}$}
}
{fig1}{(1)}\label{fig1}
\*
\0{\small
\it The Deprit angles. Here $\lis{{\bf n}}$ is the node
line $({\bf \lis i},{\bf \lis j})\cap({\bf i}_1,{\bf i}_2)$, ${\bf n}$
is the node line $({\bf i}_2,{\bf i}_3)\cap({\bf i},{\bf j})$ and
${\bf m}\equiv{\bf i}$ is the node $({\bf i},{\bf j})\cap({\bf \lis i},{\bf
\lis
j})$. The ${\bf j}$ axis is not drawn and ${\bf k} \parallel {\bf
  M}$. Colors are meant to suggest that the angles $\g,\f,\b$ lie on
different planes.}
\*

With reference to the Hamiltonian in the form Eq.\equ{e2.2}, where
$\b=0,\p$ correspond to {\it unstable} rotations around the $\bf i_1$
proper axis, the pairs of canonical variables are $(B,\b)$, $(A,\f)$ and
$(M_z,\g)$; $\sin\th =\frac{B}{A}$ while the angle $\d$ is constant. At the
``fixed points'' $B=0,\b=0,\p$ or $B=0,\b=\pm\frac\p2$ the angles $\f$ and
$\g$ rotate at constant speeds: $\dot\f=\frac{A}{I_1}$ and $\dot\g=0$ or
$\dot\f=\frac{A}{I_2}$ and $\dot\g=0$ respectively; the angle between the
angular momentum and the inertial axis $\V i_3$ is $\frac\p2$ and
$\th=0$. For the stable rotations the role of the axes $\V i_1$ and $\V
i_2$ are exchanged. Hence the coordinates are adapted to describe proper
rotations around the axis $1$ ($\b=0,\p$, unstable) of intermediate inertia
or $2$ ($\b=\pm\frac\p2$, stable) of largest inertia.

The inertia moments $I_3<I_1<I_2$ will be used to define the quantities
\be\eqalign{
J_+\defi& I_1,\kern2.3cm J_-\defi I_2,  \cr
J_{31}^{-1}\defi& I_3^{-1}-I_1^{-1},\qquad  J_{32}^{-1}\defi
I_3^{-1}-I_2^{-1},\qquad  J_{12}^{-1}\defi I_1^{-1}-I_2^{-1}\cr
\wt J_+\defi& J_{31},\kern2.1cm  \wt J_-\defi J_{32}\cr
}\Eq{e2.3}\ee
The Lyapunov coefficients at the unstable fixed point are $\pm\l_+$ real
with $\l_+\defi\frac{ A}{({\wt J_+J_{12}})^{\frac12}}$ and at the stable are
$\pm\l_-$ imaginary with $\l_-\defi \frac{i\,A}{(\wt J_-J_{12})^{\frac12}}$.

We try to use a notation valid for {\it both} the stable and the unstable
rotations: therefore a label $a=\pm1$, often abridged into $a=\pm$, is
appended to most quantities: with $a=+1$ referring to the unstable case and
$a=-1$ referring to the stable one.  Calling $U$ the total energy, define
for $a=\pm1$:
\bea
&&b^2_a\defi\frac{2U-{A^2}/{J_a}}{\wt J_{a}^{-1}},\qquad
\lis g^2_a\defi(2U-{A^2}/J_a)\wt J_a^{-1}\equiv b_a^2\,(\wt J_a^{-1})^2,
\nn\\
&&s^2_a\defi A^2\frac{ J_{12}^{-1}}{2U-A^2/J_a},\qquad
r^2_a\defi\frac{J_{12}^{-1}}{\wt J_a^{-1}},\qquad
k^2_a\defi \frac{a s^2_a -a r^2_a}{1+a s^2_a},
                                                  \Eq{e2.4}\\
&&g^2_a=\lis g^2_a\,({1+a s^2_a})
%,\quad b^2_a\equiv\frac{\lis g^2_a}{(\wt J_a^{-1})^{2}}
,\qquad
b^2_a(1+a s^2_a)=\frac{g^2_a}{(\wt J_a^{-1})^2}\nn
\eea
then $g_a$ is a characteristic inverse time scale  and, referring to
Eq.\equ{e2.2} for $a=+$ and to Eq.\equ{e2.1} for $a=-$,

\be \eqalign{
g^2_a=&A^2J_{12}^{-1}\wt J_a^{-1}(a+\frac1{s_a^2})
\cr
\dot\b =& \pm \lis g_a\,\sqrt{(1+a s^2_a\sin^2\b)(1+ ar^2_a\sin^2\b)}\cr
B=& \pm b_a\,\sqrt{\frac{1+a s^2_a\sin^2\b}{1+a r^2_a\sin^2\b}}\equiv
\pm \frac{\lis g_a}{\wt J_a^{-1}}
\sqrt{\frac{1+a s^2_a\sin^2\b}{1+a r^2_a\sin^2\b}}
\cr}                                            \Eq{e2.5}\ee
with $A,K_z\equiv A\cos\d$ constants of motion. If $a=-$ it is
$(2U-\frac{A^2}{J_-})>0$. 

Suppose first $(2U-\frac{A^2}{J_a})>0$ and small: then
\bea
&&\lis g^2_a,\qquad a\,g^2_a>0,\qquad b^2_a>0,\qquad
\cases{ r_+^2\in (0,+\infty)\cr r_-^2\in (0,1)\cr},\, 
\nn\\
&& 
s_a^2\gg1,\ \cases{0<k^2_+<1\cr k^2_->1\cr},\qquad
1+a r_a^2=\frac{\wt J_{-a}^{-1}}{\wt J_a^{-1}}=\frac{r_{a}^2}{r_{-a}^2}, 
\Eq{e2.6}\\
&&\,a s_a^2=\frac{k_a^2+a
  r_a^2}{1-k_a^2},\qquad \frac{b_a^2}{1-k_a^2}=\frac{g_a^2 \wt J_a^2}{1+a
    r_a^2}=g_a^2\wt J_+\wt J_- \nn
\eea
and motions near $B=0,\b=0$ and with energy $U$ close to $A^2/2J_a$ and
can be expressed in terms of Jacobian elliptic integrals (see
\cite{GR965} for notations).

\*\0{\it Remark:} Rotating the positions of the inertia moments in
$\wt H$ the meaning of the variables changes but the following
analysis remains essentially unchanged (up to renaming variables): it
appears in this way that the proper rotations around axis $\V i_1$ are
(linearly) unstable while the rotations around the highest and lowest
inertia axes ($\V i_2$ and $\V i_3$) are (linearly) stable.

%%%%%%%%%%%%%%%%%%%%%%%%%%%%%%%%%%%%%%%%%%%%%%%%%%%%
%%%%%%%%%%%%%%%%%%%%%%%%%%%%%%%%%%%%%%%%%%%%%%%%%%%%
%%%%%%%%%%%%%%%%%%%%%%%%%%%%%%%%%%%%%%%%%%%%%%%%%%%%
%%%%%%%%%%%%%%%%%%%%%%%%%%%%%%%%%%%%%%%%%%%%%%%%%%%%
\section{Hyperbolic and elliptic coordinates}\label{sec3}
\iniz
%%%%%%%%%%%%%%%%%%%%%%%%%%%%%%%%%%%%%%%%%%%%%%%%%%%%
%%%%%%%%%%%%%%%%%%%%%%%%%%%%%%%%%%%%%%%%%%%%%%%%%%%%
%%%%%%%%%%%%%%%%%%%%%%%%%%%%%%%%%%%%%%%%%%%%%%%%%%%%
%%%%%%%%%%%%%%%%%%%%%%%%%%%%%%%%%%%%%%%%%%%%%%%%%%%%

``Linearization'' of rigid body motions is, of course, well known to lead
to elliptic integrals, see \cite[p.144]{Wh917}. Here we rederive the
integration in a form suitable for our purposes of action-angle coordinates
construction.

Let $\sqrt{-1}=-i$ so that $g_a=\sqrt{a} |g_a|$; for $a=+$ Eq.\equ{e2.3}
describe the separatrix branches emerging from the proper rotation around
$\V i_1$ controlled by Eq.\equ{e2.2}, while for $a=-$ they describe the
motion near the stable rotation around $\V i_2$ controlled by
Eq.\equ{e2.1}.

Let $U-A^2/2J_a>0$ close to $0$; from the definitions of the Jacobian
elliptic integrals, see \cite[2.616]{GR965}, and setting

\be k'_a=\sqrt{1-k^2_a}\equiv
\sqrt{\frac{1+ar^2_a}{1+ as^2_a}},\qquad u_a=\sqrt{1+a s^2_a}\,
\lis g_a t\equiv g_a
t\Eq{e3.1}\ee
it follows (choosing the sign $+$ in the Eq.\equ{e2.5}, for instance) from
(\cite[8.153.3]{GR965} for $a=+$ and \cite[8.153.1-8]{GR965} for $a=-$
after correcting the obvious typo in \cite[8.153.8]{GR965} and by $k'_-=-i
\sqrt{k_-^2-1}$), see Appendix \ref{AppA} for details,
\bea
&&
\kern-4mm B(t)=\frac{{b_a}}{{\rm dn}(u_a,k_a)}= b_a\frac{{\rm
      cn}(-iu_a,k'_a)}{{\rm dn}(-iu_a,k'_a)},\nn\\
&&\kern-4mm \sin^2\b(t)=\frac{{\rm
      sn}^2(u_a,k_a)}{{1+a s^2_a {\rm cn}^2(u_a,k_a)}}
=\frac{-\frac{{\rm sn}^2(-iu_a,k'_a)}{{\rm cn}^2(-iu_a,k'_a)}}
{1+as_a^2\frac1{{\rm cn}^2(-iu_a,k'_a)}}
                                  \Eq{e3.2}\\
&&\kern-4mm \dot\b(t)=\frac{(1+a s_a^2)\lis g_a{\rm dn}(u_a,k_a)}{1+a
  s_a^2{\rm cn}^2(u_a,k_a)}=\lis g_a
\frac{{\rm dn}(-iu_a,k'_a) {\rm
    cn}(-iu_a,k'_a)}{1-\frac1{1+a s^2_a}{\rm sn}^2(-iu_a,k'_a) }
\nn\eea
where the last relation is deduced from the equations of motion and 
the expression for $\sin^2\b$. 

Notice that $\b(0)=0$ and if $ (2U -A^2/J_a) \to 0^+$ it is
$k^2_\pm\to1^\mp$, $a\,g^2_a\to \wt J_a^{-1} J_{12}^{-1}A^2$, $b_a\to 0^+$
and motions with this energy are ``like'' the motions close to the
separatrix or, respectively, close to equilibrium of a pendulum.

For $(2U-A^2/J_a)\,\le0$, or for $k_a>1$ hence $k'_a$ imaginary, the
above relations have to be interpreted via suitable analytic continuations.
Some of the following formulae become singular as $U\to A^2/2J_a$: the
singularity is only apparent and it will disappear from all relevant
formulae derived or used in the following.

Introduce variables useful in the following (most of them appear in the
theory of Jacobi's elliptic functions):
\be
\eqalign{
&{\bf K}(k)\defi \int_0^{\frac\p2}{(1-k^2\sin^2\th)}^{-\frac{1}2}d\th,
\cr
&x'_a\defi
a e^{-\p \frac{\KJ(k_a)}{\KJ(k'_a)}}
,\qquad
I^2 \defi 4 \wt J_+\wt J_-
\cr
&g_a^2 k'^2_a=\lis g_a^2(1+a r^2_a)\quad u_a\,=\,g_a t,\qquad  
{g}_{0,a}\,=\,|g_a|\,\frac \p{ 2\KJ(k'_a)}\cr}  
                          \Eq{e3.3}\ee
The $g_a,g_{0,a}$ depend on $k_a$ and can, and
will, be imagined as functions of $x'_a$.

\*
Let $\a_a\defi{\rm arcsin}\frac1{\sqrt{1+ar_a^2}}$$-i
  v_a=\int_0^{\a_a}\frac{d\th}{\sqrt{1-k'^2_a\sin^2\th}}$ and the integral
  can be evaluated by series, \cite[2.511.2]{GR965} and
  \cite[8.113.1]{GR965}, defining $P_m$ and $V$
\be\eqalign{ -i v_a=&\frac{2 \KJ(k'_a)}\p\a_a- \sqrt{ar_a^2}
  \sum_{m=1}^\infty {-\frac12\choose m} (-k'^2_a)^{m} P_{m}(\frac1{1+a
    r_a^2})\cr \kern3mm \defi&\frac{2 \KJ(k'_a)}\p\Big(\a_a+\sqrt{ar_a^2}
  V(\frac1{1+a r_a^2},k'^2_a)\Big) \cr}\Eq{e3.4}\ee
with $P_m(z)=\frac1{2m}(\sum_{k=0}^{m-1}
\frac{(2m-1)!!\,(m-k-1)!}{2^k\,(2m-2k-1)!! \,\ell!}  z^{\ell-k})$. This is
well defined near the proper rotations , where $k_a\to1$, as $k'_a$ tends
to $0$. Therefore, as derived in detail in Appendix \ref{AppA},
\be\g_a\equiv\G(ar_a^2,k'^2_a)\defi
e^{v_a\frac\p{2\KJ(k'_a)}}=\Big(\frac{i+\sqrt{ar_a^2}}{\sqrt{1+ar_a^2}}
\Big) e^{i \sqrt{ar_a^2}V(\frac1{1+a r_a^2},k'^2_a)}\defi 
\frac{\lis\g_a}{\sqrt{a}}
\Eq{e3.5}\ee
and  $|\g_+|=1,\ \g_-=i e^\l$ with $\l$ real and $\lis\g_-$ real,
with $\l$ real.

Introduce the new variables $p'_+,p'_-$ defined by
\be p'_\s\defi \sqrt{x'_a} e^{\s{\sqrt{a}} g_{0,a}t},\quad\s=\pm, 
\qquad {\rm hence}\quad x'_a\equiv
p'_+p'_-\,.
\Eq{e3.6}\ee
The above relations can be algebraically elaborated into the transformation
of coordinates $(B,\b)\otto (p'_+,p'_-)$, see again details in Appendix
\ref{AppA}:
\be \eqalign{ B=&R(p'_+,p'_-)\defi a\,I g_{0,a}(x'_a)
\sum_{\m=\pm1}\sum_{m=0}^\infty
  \frac{a^m\, x'^{\,m}_a\, p'_\m}{1+a(x'^{\,m}_a p'_\m)^2}\cr
  \b=&S(p'_+,p'_-) \defi a\,\sum_{n=0}^\infty\sum_{\s,\h=\pm1}
%%%%%
\frac{\h^{\frac{1+a}2} \s}i
%%%%
\,a^n\, {\rm
    arctanh}(x'^{\,n}_a p'_{\s}\lis \g_a^{\h}) \cr}
                    \Eq{e3.7} \ee
and in the new coordinates the {\it flow $t\to p'_\s e^{\sqrt{a}\s
g_{0,a}t}$ generates a solution} of the equations of motion if
$|p_\pm|$ is small and if the real parts of $p'_\pm$ are $>0$. However the
equations of motion are analytic and so are the Eq.\equ{e3.7}: hence
Eq.\equ{e3.7} yield a solution under the only condition that $|p_\pm|$ are
small, to be required for its convergence. 

Eq.\equ{e3.7} give a complete parametrization of the motions {\it but
the coordinates $p_\s$ are not canonical}: the determination of the
canonical coordinates will be discussed and done in Sec.\ref{sec4}.
\*

\0{\it Remarks:} (1) via identities relating Jacobian elliptic functions it
would be possible to avoid considering functions that are defined by
analytic continuations, for instance when the modulus $k^2_a>1$ or
$k'^2_a<0$, and reduce instead to considering only cases with the
``simple'' arguments (expressing functions by elliptic functions of other
arguments, like $h_a^2=1-k_a^{-2}, h'^2_a=k_a^{-2}$, when necessary): we
avoid this because it would hide the nice property that the stable and
unstable cases can be seen as analytic continuations of each other.  \*

\0(2) The formula for $\b$ reminds of one found by Jacobi which he
commented by saying that `` {\it inter formulas elegantissimas censeri
  debet} '', \cite[p.509]{WW927} ({\it i.e.} `` {\it it should be counted
  among the most elegant formulae} ``, see also \cite{FGG010}.
%%%%%%%%%%%%%%%%%%%%%%%%%%%%%%%%%%%%%%%%%
%%%%%%%%%%%%%%%%%%%%%%%%%%%%%%%%%%%%%%%%%
\section{Canonical coordinates construction. Jacobian}
\label{sec4}\iniz
%%%%%%%%%%%%%%%%%%%%%%%%%%%%%%%%%%%%%%%%%
%%%%%%%%%%%%%%%%%%%%%%%%%%%%%%%%%%%%%%%%%

The motions $p'_\s\to p'_\s e^{\s \sqrt{a}g_{0,a}t}$ solve the
equations of motion if the real and imaginary parts of $p'_{\pm}$ are
positive. But the equations of motion are analytic, hence the formulae of
the previous sections give solutions of the equations independently of the
sign of $p'_\s$, provided the series converge. Convergence requires
$|p'_+|,|p'_-|\ll 1$: which represents many data, in particular those in
the vicinity of the separatrix or of the stable proper rotation.

The coordinates can be called ``hyperbolic'' or ``elliptic''.  We also see
that time evolution preserves both volume elements $dBd\b$ and $dp'dq'$;
which means that the Jacobian determinant
$\frac{\partial(B,\b)}{\partial(p',q')}$ must be a function constant over
the trajectories, hence a function $D_a(x')$ of $x'\defi p'_+p'_-$ in the
two cases. Note that $D_a(x')$ has dimension of an action.

Introduce the variables $p',q'$, related to $p'_\pm$ in Eq.\equ{e3.6} by:

\be
\cases{p'=p'_+, \ q'=p'_-   & if $a=+$\cr
       p'+i\s q'=p'_\s& if $a=-$\cr},
\Eq{e4.1}\ee
to deal with real coordinates when useful; it is then possible to change
coordinates setting $p_\pm = {C_a(x'_a)}\, p'_\pm$, (or $p\defi
{C_a(x'_a)}\, p'$, $ q\defi {C_a(x'_a)}\, q'$) and choose $C_a(x')$ so that
the Jacobian determinant $D_a(x')$ for $(B,\b)\otto$ $(p_+,p_-)$ is
$\equiv1$.  A brief calculation shows that this is achieved by fixing

\be C^2_a(x')=\frac{1}{x'}\int_0^{x'} D_a(y)dy,\Eq{e4.2}\ee
which is possible for $x'$ small because, from the equations of motion it
is $D_a(0)$ finite ({\it e.g} $D_+(0)=4 I
g(0)\sin\frac{\p\a_+}{2K(k'_+)}>0$). Therefore map $(B,\b)\otto$ $(p,q)$ is
area preserving, hence {\it canonical}.  The Hamiltonian Eq.\equ{e2.1}
becomes a function $\UU_a(x_a)$ of $x_a =p_+p_-$ and the derivative of the
energy with respect to $x_a$ has to be $g_{0,a}(x'_a)$ (because the
$p_+,p_-$ are canonically conjugated to $B,\b$).  Note that $x_a$ has the
dimension of an action, while $p,q$ are, dimensionally, square roots of
action.

This allows us to find $D_a(x')$: by imposing that the equations of motion
for the $(p,q)$ canonical variables have to be the Hamilton's equations
with Hamiltonian $\UU_a(x_a)\defi U_a(x'_a)\equiv H(B, \b)$ it follows that
$\frac{d\UU_a(x_a)}{d x_a}=g_{0,a}(x'_a)$, {\it i.e.}
$\frac{d\,U_a(x'_a)}{d x'_a} \frac{d x'_a}{d x_a}=g_{0,a}$ or
$\frac{d\,U_a(x'_a)}{d x'_a}= g_{0,a}\,\frac{d}{dx'_a} (x'_a\,
C_a(x'_a)^2)2^{-\frac {1-a}2}=g_{0,a}D_a(x'_a)$ by the above expression for
$C_a(x'_a)$. The just obtained relation gives
\be D_a(x'_a)=g_{0,a}(x'_a)^{-1} \frac{d}{dx'_a}U_a(x'_a)\Eq{e4.3}\ee
which is an expression for the Jacobian
$\frac{\partial(B,\b)}{\partial(p',q')}\equiv
\frac{\partial(R,S)}{\partial(p',q')}=\frac{\partial(p,q)}{\partial(p',q')}$
(note that the Jacobian between $(B,\b)$ and $(p,q)$ is identically $1$ by
construction).

To proceed it is necessary to determine the function $C_a(x'_a)^2$.
The idea in \cite{FGG010} is to make use of the general theory of
relative cohomology classes to determine $D_a(x'_a)$ and, therefore, to
find a complete expression of the action angle variables following a
procedure employed in the theory of limit cycles of planar vector
fields, \cite{Fr996,Fr998}.

A derivation can be based directly on the equalities on symplectic forms:
\be \eqalign{
&dB\wedge d\beta=D_a(x'_a)dp'\wedge dq'= 
\frac{d}{dx'_a}[x'_a
  C_a(x'_a)^2]\,dp'\wedge dq'\cr
&=\Big(\frac{i}2\Big)^{\frac{1-a}2}\frac{d}{dx'_a}[x'_a
  C_a(x'_a)^2]\,dp'_+\wedge dp'_-\cr}\Eq{e4.4}\ee
To simplify the notation we drop for a while the label $a$ from $x'$'s.
The idea is to compute only the {\it cohomology class} of the volume forms
in the relative cohomology of the function $x'=p'_+p'_-$,
\cite{Fr996,Fr998}.

For any $2$-form $\f(p',q') dp'\wedge dq'$ with $\f(p',q') =\sum_{m,n}
f_{m,n} p'^m q'^n$ the ``{\it cohomology class}'' relative to $x'=p'_+p'_-$
is the function defined by $\psi(x')\defi$
$ \sum_{m} f_{m,m} (p'q')^m$.  Therefore an
expression for the Jacobian $D_a(x')$ is obtained by finding an expression
for the cohomology class of the form $d B\wedge d\b$ relative to
$x'=p'_+p'_-$.

Remark that Eq.\equ{e3.7} imply
\be
\eqalign{
R'(p'_+,p'_-)&dS'(p'_+,p'_-)=Ig_{0}(x')
\sum_{m,n=0}^{+\infty}\sum_{\m,\s,\h=\pm1} \frac{\h^{\frac{1+a}i}\,
\s}i\,
\cr&
\cdot\frac{a^{m+n}{x'}^m \, {x'}^n \lis\g_a^{\h}}
{(1+a({x'}^m p'_\m)^2)\,(1-({x'}^n p'_{\s}\g_a^{\h})^2)}\,p_\m dp'_{\s}+\V0
\cr
}\Eq{e4.5}\ee
where $\V0$ is a form with $0$ cohomology ({\it i.e} the first term is
$RdS$ ``modulo $d(p'_+p'_-)$'' because the cohomology of $d
\Big(G(p'_+p'_-) d(p'_+ p'_-)\Big)$ is, for all $G$'s, $0$).

The terms with $p_+dp_+$ and $p_-dp_-$ do not contribute to the cohomology
and, up to terms (denoted $\bf 0$) not contributing to the cohomology, the
form in Eq.\equ{e4.4} becomes (as only terms with $\m=-\s$ contribute)
\be
\eqalign{
R'(p'_+,p'_-)&dS'(p'_+.p'_-)= Ig_{0}(x')\frac{1}i
\sum_{m,n=0}^{+\infty}\sum_{\h,\s=\pm1}  x'
\frac{\s dp'_{\s}}{p'_{\s}}\cr
&\cdot
\frac{a^{m+n}{x'}^{m+n}\,\h^{\frac{1+a}2}\,
\lis \g_a^{\h}}
{(1+a({x'}^m p'_{-\s})^2)\,(1-({x'}^n\lis \g^{\h}p'_{\s})^2)}+\V0
\cr}
                     \Eq{e4.6}\ee
If $RdS= x'F(p'_+,p'_-) \frac12(\frac{d p_-'}{p'_-}-\frac{d p_+'}{p_+'})
+\V 0$ and $F(p'_+,p'_-)=\sum_{h,k=0}^\infty F_{h,k}p'^h_+,p'^k_-$, only
the terms with $k=h$ contribute and we can take
\be\eqalign{
 F=&Ig_{0}(x')\frac{1}{i}
\sum_{m,n=0}^\infty\sum_{\h,\s=\pm1}
\frac{a^{m+n}{x'}^{m+n+1}\,\h^{\frac{1+a}2}\lis \g^{\h}_a}
{(1+a({x'}^m p'_{\s})^2)\,(1-({x'}^n\lis \g^{\h}_ap'_{-\s})^2)}\cr}
           \Eq{e4.7}\ee
hence the cohomology with respect to $p_+p_-$ is that of
$(\frac{dp_+}{p_+}-\frac{dp_-}{p_-})$ times
\be\kern-4mm\eqalign{
& \frac{2Ig_{0}(x')}{i}\sum_{\h=\pm1}\sum_{m,n,h} x' \h^{\frac{1+a}2}
a^{m+n}x'^{m+n}\lis \g^\h_a
(x'^{m+n+1}\lis \g^\h_a)^{2h}
\frac{(-a)^h}2
\cr
&=Ig_{0}(x')\,
 \,x'\sum_{\h=\pm}\frac{\h^{\frac{1+a}2} }i
\sum_{\ell=0}^\infty \frac{(\ell+1)\,a^\ell\,x'^\ell\,\lis \g^\h_a}{
1+a(x'^{(\ell+1)}\,\lis \g^\h_a)^2}\cr} \Eq{e4.8}
\ee
which contributes 
\be-\frac{d}{dx'}\Big(2I \,x'\,g_{0}(x')\,
\sum_{\h=\pm1}\frac{\h^{\frac{1+a}2}}i\sum_{\ell=0}^\infty
 \frac{(\ell+1){x'}^{\ell}\,a^\ell\,\lis \g^\h_a}{1+a
   ({x'}^{(\ell+1)}\lis \g^\h_a)^2}\Big)\,dp'_+\wedge dp'_-
\Eq{e4.9}\ee
because the cohomology of $d(x'f(x')
(\frac{dp'_+}{p'_+}-\frac{dp'_-}{p'_-}))$ is
$-2\frac{d}{dx'}(x'f(x'))dp'_+\wedge dp'_-$.  Therefore, by Eq.\equ{e4.2}
and by $dp'_+\wedge dp'_-=(\frac2{i})^{\frac{1-a}2}dp'\wedge dq'$, see
Eq.\equ{e3.5}:
\be\eqalign{
& 
{C_a(x')^2}=\,2\,I\,g_{0,a}(x')\,\sum_{\h=\pm1}
\Big(\frac{\h}{ i}\Big) ^{\frac{1+a}2}
\sum_{\ell=0}^\infty
\frac{(\ell+1)a^\ell{x'}^{\ell}\lis 
\g^\h_\a}{1+ a({x'}^{(\ell+1)}\lis \g^\h_\a)^2}
\cr
& \equiv2\, I\,g_{0,a}(x')\,\sum_{\h=\pm} \frac{\h}{i}
\sum_{\ell=0}^\infty
\frac{(\ell+1)a^\ell{x'}^{\ell} 
\G(a r_a^2, k'^2_a)^\h}{1+ ({x'}^{(\ell+1)}\G(a r_a^2, k'^2_a)^\h)^2}\cr
&\defi
2 A r_a C(ax'_a, ar_a^2)^2
\cr}\Eq{e4.10}\ee
Summarizing we set, with $p'_\pm$ defined in Eq.\equ{e4.1},\equ{e3.6},
\be
\eqalign{
&p=\sqrt{2A r_a}\, p'\, C(ax'_a,a r_a^2),\quad q=\sqrt{2A r_a}\, q'\, 
 C(ax'_a,a r_a^2),
\cr
&x'=f(p'_+p'_-), \ \frac{p_+p_-}{2A r_a}=x_a=x'_a C^2(ax'_a,ar_a^2)\cr}
\Eq{e4.11}\ee
where $p'_+p'_-=p'q'$ if $a=+$ and $p'_+p'_-=(p'^2+q'^2)$ if $a=-$.

%%%%%%%%%%%%%%%%%%%%%%%%%%%%%%%%%%%%%%%%%%%%%%%
%%%%%%%%%%%%%%%%%%%%%%%%%%%%%%%%%%%%%%%%%%%%%%%
\section{Evaluation of Jacobians and normal forms:}
\iniz\label{sec5}
%%%%%%%%%%%%%%%%%%%%%%%%%%%%%%%%%%%%%%%%%%%%%%%
%%%%%%%%%%%%%%%%%%%%%%%%%%%%%%%%%%%%%%%%%%%%%%%

It it possible to evaluate $C_a^2$ and the normal form for the energy $U=H$
in the canonical coordinates $(p_+,p_-)$ up to large orders in $x_a$.  This
is done simply by deriving power series expansions, in $x'_a$ first and
then $x'_a$ in terms of $x_a$, for the functions in Eq. \equ{e5.3}.

Consistently with Eq.\equ{e4.11} $C_a^2$ can be expanded into a power
series of $x$ or $x'$ whose coefficients are functions of the parameter
$r_a^2$. There is a simple relation between the cases $a=\pm$ and the
coefficients of the expansions in $x$ can be computed exactly with a finite
computational algorithm up to any prefixed order: this is described below.

The analysis of the Jacobian $C_a(x,r)^2$ can be based on the 
computable expressions of $\lis \g_a$ in Eq.\equ{e3.5}.

Normal forms express $\wt H_a\defi \frac{2H -A^2 J_a^{-1}}{a
  J_{12}^{-1}}$.  From Eq.\equ{e2.1},\equ{e2.2} we derive $A^2 s^{-2}_a$
  (see Eq.\equ{e2.6}) in terms of $k^2_a$ and, by Eq.\equ{e3.3}, $x'_a$ in
  terms of $k^2_a$:
\be\eqalign{
 x'_a=&ae^{-\p \frac{{\bf K}(k_a)}{{\bf K}(k'_a)}},\qquad a=\pm
\cr}\Eq{e5.1}\ee
which can be inverted in terms of the functions
$F(z)=\frac{2\sum_{n=1}^\infty z^{(2n-1)^2}}
{1+2\sum_{n=0}^\infty z^{(2n)^2}}$ as
\be
k_a(x')^2=
\Big(\frac{1-F(ax')}{1+F(ax')}\Big)^{4}\defi W(ax'),
\qquad a=\pm1\Eq{e5.2}\ee
as shown in \cite[8.198]{GR965}.
Therefore the above relations give explicitly the functions $\wt H_a$ in
terms of $x'$, hence they determine the functions $U_a(x')$.

To express $\wt H_a$ in terms of $x$ we have to determine the Jacobian
function and invert the relation $x=x' \wt C_a^2(x')$ in the two cases.
This is done by remarking that $g_{0,a}=
\sqrt{a}\,g_a\frac{\p}{2\KJ(k'_a)}$ and $U_a(x')$ can be written as
\bea
&&\frac{g_{0,a}(x')}{\sqrt{a}\,g_a(x')}=\frac\p{2\KJ(k'_a)}
\quad ag_a(x')^2=
\frac{A^2r_{-a}^2} {k_a^2+ar_a^2}\frac1{J_{12}^2},
\nn\\
&&\wt H(x')\defi\frac{2U_a(x')-A^2/ J_a}{aJ_{12}^{-1}}={A^2}
    \frac1{as_a^2}=A^2\frac{
      (1-k_a^2)}{k_a^2+a r_a^2}\nn\\
&&C^2(x',r_a^2)= 2A r_a
(W(ax')+a    r_a^2)^{-\frac12}
\frac{\p}{2\KJ(k'_a)}
                 \Eq{e5.3}\\
&&\kern9mm\cdot 2\,\sum_{h=0}^\infty\sum_{\h=\pm}
\frac{\h}{i}
\frac{(h+1)(ax')^h \G(ar_a^2,k'^2_a)^\h}{1+((ax')^{h+1}
\G(ar_a^2,k'^2_a)^\h)^2}\nn
\eea
where the relations $r_{-a}^2= \frac{r_a^2}{1+ar_a^2}$, $\frac{I
  r_{-a}}{J_{12}}=2r_a$, following from Eq.\equ{e2.6} or \equ{e2.3}, have
been used and the function $\G$ is defined in Eq.\equ{e3.5}.

This is remarkable because the summation over $\h=\pm$ is analytic in $a
r^2_a$ in spite of the fact that each addend has a branch point in $\sqrt{a
  r_a^2}$ at $r_a=0$. Therefore Eq.\equ{e5.3} shows that the two cases
$a=\pm$ are described by the same functions ({\it of course evaluated at
  completely different points}) and the power series for $C$ and $U$ can be
computed setting $a=+$ and then the case $a=-$ is obtained by replacing $x$
by $-x$ and $r^2$ by $-r^2$, $A r_+$ by $A r_-$.  The simple relation
between the unstable and stable cases implies that that we only need to
study one of the two cases: we shall only make computations in the
hyperbolic case $a=+$.

At this point the Jacobian and the normal forms can be computed as formal
series in $x$ after solving the implicit function problem $x=x'
C^2(x',r^2)$ in the form $x'=x \lis
C^2({x},r^2)$ (in a power series in ${x}$).

An algebraic numerical evaluation for $a=+$ (unstable case with, for simplicity,
$r=r_+,x'_+=x'$) yields results that for $n\le 7$, if $\x\equiv
\frac{x'}{(1+r^2)}$, are
\bea
&&x\,C(x,r^2)^2=4 \x -
 24 (-1 + r^2) \x^2 + 32 (3 
- 14 r^2 + 3 r^4) \x^3
\nn\\
&&- 16 (-1 + r^2) (19 - 242 r^2 + 19 r^4) \x^4\nn\\
&& + 
24 (35 - 1140 r^2 + 3026 r^4 - 1140 r^6 +
      35 r^8) \x^5 \Eq{e5.4}\\
&&- 192 (-1 + r^2) (11 - 740 r^2 + 3426 r^4 -
      740 r^6 + 11 r^8) \x^6\nn\\
&& + 64 (77 - 10206 r^2 + 103635 r^4 - 211460 r^6 +
      103635 r^8\nn\\
&&\kern4mm - 10206 r^{10} + 77 r^{12}) \x^7+\ldots\nn
\eea
Finally the polynomials in $r^2$ in Eq.\equ{e5.4}, coefficients of order
$n>1$ for the power expansion in $\x^n$, {\it can be conjectured to have 
  all roots real}: this has been numerically checked for $n\le15$.

The normal forms of the Hamiltonians ({\it i.e.} their expression in the
variables $p,q$) are derived by expressing $s^2$ in Eq.\equ{e2.1} in terms
of $r^2$ and $k^2$ and then expanding $k^2$ in powers of $x'$ using $x'
=e^{\p \frac{\KJ(k)}{\KJ(k')}}$ in Eq.\equ{e5.1}, via \cite[8.198]{GR965}
(notice that in our hyperbolic case $x'$ is the $q$ of the quoted tables
with $k$ and $k'$ exchanged and finally computing $x'=x\,{\lis C}^2(x,r^2)$
by inverting the Jacobian relation $x=x'\,C^2(x',r^2)$.  This can be
implemented algebraically on a computer.

With the definitions in Eq.\equ{e2.3} $\wt
H_a\defi \frac{2H -A^2J_a^{-1}}{a\,J_{12}^{-1}}$, in the unstable case
($a=+1$) or in the stable ($a=-1$) case, can be expressed
 as function of $x=\frac{pq}{2Ar_+}
$ or, respectively, of $x=\frac{p^2+q^2}{2Ar_-}$ obtained by solving $x=x'
C^2(x',r^2)$ as $x'=x\X(x,r^2)$ and replacing $x'$ by $\X(x,r^2)$ in $U(x')$
in Eq.\equ{e5.3}(notice that ${J_{12}^{-1}}={r_a^2}{\wt J_a^{-}}$). The
resulting functions are normal forms for the Hamiltonian since $p,q$ are
conjugated variables. Hence

\be H(x_a)=\frac{A^2}{2J_a}+ \frac{ A^2 \,a r_a^2}{2\wt J_a} 
\,\HH(ax_a,ar_a^2)
\Eq{e5.5}\ee
for $a=\pm1$ where $\HH(x,r^2)\defi \frac1{a
  s_a^2}=\frac{1-W(\X(x_a,ar_a^2))}{W(\X(x_a,ar_a^2))+ar_a^2}$, where
$W(x)$ is in Eq.\equ{e5.2}. One
obtains
\bea%\eqalign{
&&
\HH(x,r^2)=4 x - 2 (-1 + r^2)x^2 - (1 + r^2)^2 x^3\nn\\
&& -
\frac54 (-1 + r^2) (1 + r^2)^2 x^4
-
 \frac3{16} (1 + r^2)^2 (11 - 10 r^2 + 11 r^4) x^5
\nn\\
&& -
 \frac7{16} (-1 + r^2) (1 + r^2)^2 (3 - 4 r + 3 r^2) (3 + 4 r +
    3 r^2) x^6\Eq{e5.6} \\
&&-
 \frac1{64} (1 + r^2)^2 (527 - 332 r^2 + 330 r^4 - 332 r^6 + 527 r^8) x^7
\nn\\
&& -
 \frac{9}{512} (-1 + r^2) (1 + r^2)^2 (1043 + 548 r^2 + 1058 r^4
\nn\\
&&\kern4mm +
    548 r^6 + 1043 r^8) x^8+\ldots
\nn
\eea
It is remarkable that {\it the polynomials in $r$ appearing as coefficients
  of order $2\le n\le 15$ in the power series expansion in $x$ of the 
  normal form have, for all $n$'s, just roots of unity} as it appears by
evaluating them numerically.

%%%%%%%%%%%%%%%%%%%%%%%%%%%%%%%%%%%%%%%%%%%%%%%
\section{Normal form without determination of action-angle variables}
\iniz\label{sec6}
%%%%%%%%%%%%%%%%%%%%%%%%%%%%%%%%%%%%%%%%%%%%%%%

The function $\HH(x)$ being the normal form of the Hamiltonian in
action-angle variables is unique, while the action-angle variables
themselves are not unique. Therefore it makes sense to see if the normal
form itself could be derived without actually determining the angles
(notice that the actions being a function of the energy are uniquely
determined). Here we show that this can be achieved quite directly without
having to deal with elliptic integrals: while in the elliptic case, for
instance, the classical prescription, \cite[Sec.50]{Ar989}, would be to
study the Hamitonian as a function of the integral $\oint
B \frac{d\b}{2\p}$
with $B$ defined in Eq.\equ{e2.5}).

As it is well known the full determination of the action-angle coordinates
is much more difficult than the determination of the normal form of the
Hamiltonian, see for instance the ``conjecture'' in \cite{GM973} and its
proof in \cite{Fr988}. Therefore a direct determination of the normal form
is worth investigating: and as it will appear here, the construction
yields unexpected insights into our problem.

Consider first the normal form near the stable rotation.  In contrast with our
first method, we do not start with the Jacobi's coordinates in which the
motion is linear.

The Hamiltonian value $U$ will be written as (see Sec.\ref{sec2}, $a=-$): 
\be\frac{2U-A^2/I_2}{J_{32}^{-1}}=
B^2(1-r^2_a \sin^2\b) +r^2_a A^2\sin^2\b
\Eq{e6.1}
\ee
For simplicity of notation the label $a=-$ will be omitted.
Introducing $X=B\sqrt{1-r^2(\sin\b)^2}$, $Y={rA}\sin\b$ the Hamiltonian
becomes $U=\frac1{2J_{32}}(X^2+Y^2)+\frac{A^2}{2I_2}$ and the form $\o=d
B\wedge d\b$ is, if $X_\pm\defi \frac{X\pm i Y}{\sqrt2}$,
\bea
&& dB\wedge d\b=\frac1
{\sqrt{{(1-r^{2}(\frac{Y}{rA})^2)(1-(\frac{Y}{rA})^2)}}} \frac{dX\wedge
dY}{Ar}
\nn\\
&&=\sum_{n=0}^\infty\sum_{h+k=n} {-\frac12\choose h} {-\frac12\choose
  k}(-1)^{k+h} r^{2h}\Big(\frac{X_+-X_-}{i\sqrt2 A
  r}\Big)^{2(k+h)}\frac{dX\wedge dY}{A r} 
\nn\\
&&=\sum_{n=0}^\infty P_n(r^2) \Big(\frac{X_+ X_-}{2 A^2 r^2}\Big)^n 
\frac{dX\wedge
  dY}{A r}+ {\bf 0},\qquad{\rm where}\ \    P_n(r^2)             \Eq{e6.2} 
\\
&&=(-1)^n{2n\choose n} \sum_{h+k=n} {-\frac12\choose h}
{-\frac12\choose k} r^{2h}={2n\choose n}\sum_{h+k=n}{2h\choose h}{2k\choose
  k}r^{2h}
%\nn\\
%&&={2n\choose n}\sum_{h+k=n}{2h\choose h}{2k\choose  k}r^{2h}
\nn\eea
and ${\bf 0}$ is a form with  $0$--cohomology relative to $X_+X_-$. 

%%%@@@@@@@
Therefore, see \cite[proposition 1]{Fr978},\cite{Gu981}, or Theorems 1.1,
1.2 in \cite{Fr988b}, there is a change of coordinates $X,Y\to X',Y'$
which, if $H'=\frac{(X'^2+Y'^2)}{(2Ar)^2}$ and
$H=\frac{(X^2+Y^2)}{(2Ar)^2}$, keeps $H$ equal to $H'$ and makes $\o$
proportional to $dX'\wedge dY'$:

\be\eqalign{
\o=& 
\sum_{n=0}^\infty P_n(r^2) 
H'^n \frac{dX'\wedge   dY'}{Ar}
\defi D(H')\,\frac{dX'\wedge   dY'}{Ar}\cr}
\Eq{e6.3}\ee
Let $\wh p=\frac{X'}{\sqrt{2Ar}} \wh C(H'), \,\wh q=\frac{Y'}{\sqrt{2Ar} }
\wh C(H')$ with $\wh C$ chosen, see also Sec.\ref{sec4}, so that $dB\wedge
d\b=d \wh p\wedge d\wh q$, {\it i.e.} as $\wh C(z)^2+z\frac{d\,\wh
  C(z)^2}{dz}=D(z)$ or

\be{ \wh C(H)^2=
\sum_{n=0}^\infty \frac{P_n(r^2)}{(n+1)}
H^n},\Eq{e6.4}\ee
For instance, up to $O(z^7)$:
\be \eqalign{
\textstyle &z\,\wh C(z)^2=\textstyle 
z 
- \frac{1}{2 }(-1 - r^2) z^2 
+ \frac{1}{4 }(3 + 2 r^2 + 3 r^4) z^3 \cr&
- \frac{5}{16 }(-1-r^2)(5 - 2 r^2 + 5 r^4) z^4\cr
& 
+ \frac{7}{64 }(35 + 20 r^2 + 18 r^4 + 20 r^6 + 35 r^8) z^5 \cr&
- \frac{21}{128 }(-1-r^2)(63 - 28 r^2 + 58 r^4 - 28 r^6 + 63 r^8) z^6 \cr&
+\frac{33}{256 }(231 + 126 r^2 + 105 r^4 + 100 r^6 + 105 r^8 + 126 r^{10} 
         + 231 r^{12}) z^7 \cr
}\Eq{e6.5}\ee
and a normal form of the Hamiltonian is 

\be U(\wh p,\wh q)=
\frac{A^2}{2I_2}+\frac{A^2 r^2}{2 J_{32}} {\cal H}(x,r^2)\Eq{e6.6}\ee
where ${\cal H}(x,r^2)$ is $4$ times the inverse function to $x\,=\,z\, \wh
C (z)^2$ and $x=\frac{\wh p^2+\wh q^2}{2Ar}$ (with $ r=r_-$): this gives a
normal form, because the variables $\wh p,\wh q$ are canonically conjugated
to $B,\b$. For instance (and in agreement with Eq.\equ{e5.6}):
\be\eqalign{
&\textstyle{\cal H}(x)=
4 x -2 (1 +  r^2) x^2 - (-1 + r^2)^2 x^3 
-\frac{5(1 + r^2) (-1 + r^2)^2
}4  x^4 \cr
&\textstyle-\frac{3(-1 + r^2)^2
}{16}  (11 + 10 r^2 + 11 r^4) x^5 
-\frac{7(-1 + r^2)^2(1+r^2) 
}{16} (9-2 r^2+9 r^4) x^6\cr
&\textstyle-\frac{ (-1 + r^2)^2
}{64} (527 + 332 r^2 + 330 r^4 
+ 332 r^6 + 527 r^8)
x^7\cr
&
\textstyle-\frac{9(1 + r^2) (-1+r^2)^2}{512} (1043 - 548 r^2 
+ 1058 r^4
-548 r^6 + 1043 r^8) x^8 +\ldots\cr} \Eq{e6.7}
\ee
It should be noted that the above analysis does not determine the canonical
map into action-angle variables of the initial coordinates and, in
particular, a further change of coordinates which transforms the plane $\wh
p,\wh q$ into itself by rigidly rotating all circles centered at the origin
by a radius dependent (and even $r$ dependent) angle will lead to a normal
form. For instance calling $\z=p+iq$ the change of variables $\z'=\z e^{i
  |\z| Q(r^2)}$ is a canonical map, for any choice of $Q(r^2)$, and it
keeps the system in normal form.

Since the Hamiltonian is uniquely cast in normal form, the polynomials
$P_n(r^2)$, coefficients of the  expansion of ${\cal H}$ in powers of $x$,
have an intrinsic significance. It is therefore interesting that they seem
to enjoy the property of having all zeros on the unit circle:
we have checked this for $1<n<15$: are really the {\it zeros  
  on the unit circle for all $n$?}. 

The location on the unit circle of the roots of our polynomials is certainly
related to  the property that the polynomials

\be\lis P_n(r^2)\defi \sum_{h+k=n} {-\frac12\choose h} {-\frac12\choose
k}\, r^{2h}\equiv
\frac1{2^{n}}\sum_{h+k=n} {2h \choose h} {2k \choose
k}\, r^{2h}
\Eq{e6.8}\ee
appearing in the above expression for $\wh C^2$ have all the roots on the
unit circle.

The latter is a special case of the general result on polynomials, that we
call ``polynomials with {\it symmetric, positive and monotonic
  coefficients}'', of the form $Q(z)=\sum_{h=0}^{n} F_h z^h$, with
$F_h=F_{n-h}$ and $F_{h-1}>F_h>0$ for $0\le h\le [\frac{n+2}2]$: \*

\0{\bf Theorem: \rm (Chen, 1995) \it Polynomials with symmetric, positive
  and monotonic coefficients have all roots on the unit circle.}  \*

The theorem can be found in \cite{Ch995}, and it is described, for
completeness in Appendix \ref{AppB}.

This is analogous to the conjecture formulated after
Eq.\equ{e5.4}, and it is an algebraic property which could, we think, also
be seen as follows. Consider the polynomials

\be Q(z^2)=z^n\sum_{\s_1,\ldots,\s_n=\pm1} e^{\sum_{i,j}
  J_{ij}(\frac{\s_i+\s_j}2)^2}
z^{\sum_i\s_i}\Eq{e6.9}\ee
with $J_{ij}>0$. Then the coefficients $Q_h$ of $z^{2h}$ in $Q$ have the
form of a sum of ${n\choose h}$ exponentials of sums of some of the
$J_{ij}$: the sums define quantities $a_h=a_{n-h}\defi e^{p(h)}$. By
adjusting the constants $J_{ij}$ the parameters $p(h-1)-p(h)$ for $0< h\le
[\frac{n+2}2]$ can independently take any value $>0$ so that $
e^{p(h)-p(n)}=\frac{F_h}{F_n}$ can be achieved by suitable choices of the
$J_{ij}>0$. Therefore $Q$ would have all zeros in the unit circle by
Lee-Yang's theorem, \cite{Ru971}, achieving a alternative
proof of the above theorem.

The latter is a conjecture that can be checked rigorously for $n\le 5$ and
we found some preliminary numerical evidence that it should also work at
least for $n\le 8$. The proof for $n=4$ emerged from a discussion with
A. Giuliani; he also has a proof solving the case $n=5$ and
gave us the reference to Chen's theorem.

The polynomials in $r^2$ appearing as coefficients of $x^k$ in the normal
form Eq.\equ{e6.7} are sums of products of polynomials $P_{n_j}$ of degrees
which add up to $2(k-1)=\sum_j n_j$: an expression for the inverse function
to $z\wh C(z)^2$ can be found in terms of a ``tree expansion'' (quickly
sketched in Appendix \ref{AppC}).  \*

Sums of products of symmetric polynomials with symmetrically decreasing
positive coefficients is a polynomial with the same symmetry but it will
not be, in general, decreasing. Hence the above theorem on the zeros of
polynomials does not allow us any conclusion on the location of the zeros
of the coefficients of the normal form: {\it however we see empirically
  that the zeros are apparently still on the unit circle} (up to $n=15$ at
least).

This is a special property of our polynomials $P_n$ because, in general,
the inverse function to the map $x=z+\sum_{n=1}^\infty P_n(w)z^{n+1}$ with
$P_n(w)$ a degree $n$ symmetric decreasing polynomial with positive
coefficients (hence with all zeros on the unit circle) can be written
$z=x+\sum_{n=1}^\infty Q_n(w) x^{n+1}$ but the $Q_n$ {\it do not have}, in
general, all zeros on the unit circle.  \*

The analysis of the hyperbolic case is essentially identical: introducing
the variable $X_\pm =B\sqrt{1+r^2\sin^2\b}\pm A r\sin\b$ the form $d
B\wedge d\b$ is

\be\eqalign{ 
\frac{d X_+\wedge d X_-}{2Ar\sqrt{(1-\sin^2\b)(1+r^2
    \sin^2\b)}}=\frac1{2Ar}\sum_{n=0}^\infty 
\Big(\frac {-X_+X_-}{(2 A r)^2}\Big)^n
P_n(-r^2)
\cr}\Eq{e6.10}
\ee
and the normal form is related to the elliptic case and given by
$U(p,q)=\frac{A^2}{2I_2}-\frac{x A^2 r^2}{2 J_{31}} 
{{\cal H}}(-x,-r^2)$.
where $x=\frac{pq}{2Ar}$ and $r=r_+$. 
\*

\0{\it Remarks:} (1) The normal form in this section coincides with the one
in Sec.\ref{sec5} because of the normal form {\it uniqueness}. The
agreement continues up to order $15$: however the above analysis shows that
this {\it must hold} to all orders: {\it i.e.} $\frac{H}{4A^2r_a^2}$
coincides with $x_a$ and our construction in Sec.\ref{sec2}-\ref{sec5}
determines explicitly one among the changes ({\it which are not unique}) of
variables $(X,Y)\otto(X',Y')$ realizing the cohomological equivalence of
the symplectic forms $dX'\wedge dY'$ and $dX\wedge dY$.  \*
\0(2) It would be interesting to find a mechanical interpretation of the
properties of the roots of the polynomials $P_n$.

%%%%%%%%%%%%%%%%%%%%%%%%%%%%%%%%%%%%%%%%%%%%%%%%%%%%%%
%%%%%%%%%%%%%%%%%%%%%%%%%%%%%%%%%%%%%%%%%%
\section{Action-angles for the ``momentum'' degrees of freedom 
\texorpdfstring{$(A,\f)$}{}}
\iniz\label{sec7}
%\renewcommand{\theequation}{B%\arabic{section}.
%\arabic{equation}}\label{AppC}
%%%%%%%%%%%%%%%%%%%%%%%%%%%%%%%%%%%%%%%%%%%%%%%
%%%%%%%%%%%%%%%%%%%%%%%%%%%%%%%%%%%%%%%%%%%%%%%

Consider the elliptic motions, {\it i.e.} rotations close to the proper
rotation about the axis ${\bf i}_2$. Here we compute the second pair of
canonical variables corresponding to the second degree of freedom and
described canonically by $(A,\f)$.
Setting $\sn'\defi \sn(-iu h'^{-1},h)$,
$\cn'\defi \cn(-iu h'^{-1},h), \dn'\defi$ $
\dn(-iu h'^{-1},h)$,\\ Eq.\equ{e3.2} implies
\bea
&&  \sin^2\b=h'^2\frac{\sn'^2/\dn'^2}{1-s^2 \dn'^2/\cn'^2}
=\frac1{k^2} \frac{\sn'^2}{(1-h^2\sn'^2)(1-s^2\dn'^2/\cn'^2)}
                    \Eq{e7.1})\\
&&=\frac1{k^2} \frac{\sn'^2 (1-\sn'^2)}
{(1-h^2\sn'^2)(1-\sn'^2 -s^2(1-s^2 \sn'^2))}\nn\\
&& =\frac1{k^2} \frac{\sn'^2 (1-\sn'^2)}{(1-h^2 \sn'^2)(1-s^2)(1-\sn'^2)}=
\frac1{k^2 (1-s^2)} \frac{\sn'^2}{\dn'^2}
=
\frac{1}{r^2-s^2} \frac{\sn'^2}{\dn'^2}\nn\eea
and this yields
\be\dot\f=\frac{A}{I_2}+\frac{A}{I_2}(\frac{I_2}{I_1}-1)\sin^2\b\Eq{e7.2}\ee
therefore, \cite[8.146.11]{GR965}, the expression of $\f(t)$ and of the
uniformly rotating angle $\ps(t)$ with angular velocity $\o_0(A,x)$
(with $x=x(x')^2 x'$) are

\be\eqalign{
&\ps=\f-(\frac{g_{0,a}}g)^2\frac1{1-r^2}\sum_{\s,\s'=\pm1}\sum_{n\s+m\s'\ne0}
(-1)^{n+m+1}\cr
&\kern5mm\cdot\s\s'\frac{x'^{n+m-1}}{(1+x'^{2n-1})(1+x'^{2m-1})}
e^{i(\s(2n-1)+\s'(2m-1))g_{0,a}t}\cr
&\o_0(A,x')=\frac{A}{I_2}\,\Big(1+
\frac{2({I_2I^{-1}_1}-1)}{1-r^2}\frac
{g_{0,a}^2}{g^2}
\sum_{n=1}^\infty \frac{x'^{2n-1}}{(1+x'^{2n-1})^2}\Big)\cr
}                        \Eq{e7.3}\ee
with $n,m\ge1$.

Since the generating function of the canonical map integrating the motions
has the form $S(A',x,\b)=A'\f+S_0(A',x,\b)$, \cite[Problem 4.11.5]{Ga983},
it is $A'=A$ and the angle conjugated to $A$ must differ by a constant from
$\ps$ and therefore it can be taken equal to $\ps$, {\it i.e}:

\be\eqalign{
&\ps=\f-(\frac{g_{0,a}}g)^2\frac1{1-r^2}\sum_{\s,\s'=\pm1}\sum_{n\s+m\s'\ne0}
(-1)^{n+m+1}\cr
&\kern5mm\cdot\s\s'\frac{z_\s^{2n-1}
z_{\s'}^{2m-1}}{(1+x'^{2n-1})(1+x'^{2m-1})}
\cr}\Eq{e7.4}\ee
with $z_\s=(p'+i\s q')$, $x'=p'^2+q'^2$ and $p'=p a(x')^{-1},q'=q a(x')^{-1}$.

This completes the construction of the elliptic canonical coordinates
conjugated to $(B,\b,A,\f)$. The hyperbolic case, {\it i.e.} rotations close
to the proper rotation around ${\bf i}_1$, is treated in the same way.

The case of the stable rotations close to the stable proper rotation around
$\V i_3$ is also interesting and could be treated in a similar way leading
to a complete parametrization of the motion along the unstable manifolds
emerging from the rotations around the axis $\V i_3$ which correspond
$A=B$. However the coordinates $(B,\b)$ have a singularity on such
rotations (notice that the $\b$ coordinates disappear from $H$): hence new
$(B,\b)$ coordinates have to be introduced. The simplest is to use the
symmetry on the permutations of the inertia moments and make use of the
equivalent Hamiltonian $H(B,\b)=\frac12
\frac{B^2}{I_2}+(\frac{\sin^2\b}{2I_3} +\frac{\cos^2\b}{2I_1})(A^2-B^2)$ in
which the coordinates $B,\b$ have a suitably different meaning but can be
used to describe the motions close to proper rotations around the axes $\V
i_3$ and $\V i_1$ (respectively stable and unstable). The analysis is then
a repetition of the case $a=+$ with the Hamiltonian Eq.\equ{e2.2}.

%%%%%%%%%%%%%%%%%%%%%%%%%%%%%%%%%%%%%%%%%%%%%%%%%%%%%%
%%%%%%%%%%%%%%%%%%%%%%%%%%%%%%%%%%%%%%%%%%
\section{Pendulum and ellipsoid of revolution}
\iniz\label{sec8}
%%%%%%%%%%%%%%%%%%%%%%%%%%%%%%%%%%%%%%%%%%%%%%%
%%%%%%%%%%%%%%%%%%%%%%%%%%%%%%%%%%%%%%%%%%%%%%%

The method of Sec.\ref{sec6} can be applied to the case of the pendulum
($H=\frac{B^2}{2I}+Ig^2(1-\cos\b)$, \cite{FGG010}) or to the geodesic
motion on a revolution ellipsoid
($H=\frac{B^2}{2(b^2\sin^2\th+a^2\cos^2\th)}+ \frac{A^2}{2 a^2\sin^2\th}$,
\cite[(4.12.6)]{Ga983}). 

The first case is considered in an attempt at proving the sign property of
the normal form coefficients suggested in \cite{FGG010}.  In the second
case it is of interest because the normal form has the form of a series
with coefficients polynomials in the ratio $r^2=\frac{a^2}{b^2}$ pf the
equatorial axis $a$ to the polar axis $b$ and in this case the zeros of the
polynomials do not seem to have special properties (in particular they are
not located on the unit circle).

The normal form at the stable equilibrium of the pendulum is easily
evaluated (details are skipped) to be $8I g^2\HH(\frac{\wt p^2+\wt q^2}{16I
  g})$ if $\HH(x)$ is the inverse function to $x=z\wt C(z)^2$ with:

\be\wt C^2(z)\defi\sum_{n=0}^\infty{2n\choose n} {-\frac12\choose n}
 \frac{(-z)^n}{n+1}
\equiv
\sum_{n=0}^\infty{2n\choose n}^2 \frac{z^n}{4^n(n+1)}
%\sum_{n=0}^\infty{2n\choose n} \frac{(2n-1)!!}{2^n(n+1)!}z^n
\Eq{e8.1}\ee
(which is $F(\frac12,\frac12;2;4z)$,
see \cite[9.100]{GR965}). And to order $O(x^8)$ the $\HH$ is:
% \cite[5.7(f)]{PWZ997}
%
\be 
\HH(x)=x-\frac12x^2-\frac14x^3-\frac{5}{16}x^4
-\frac{33}{64}x^5-\frac{63}{64}x^6
-\frac{527}{256}x^7-\frac{9387}{2048}x^8-...\Eq{e8.2}\ee
This is consistent with the first few terms computed in
\cite[(C.8)]{FGG010}: due to different meaning of the symbols the relation
between the function $\HH(x)$ above and $W(x)$ in the quoted reference is
$W(x)=\frac14\HH(4x)$. However constructing $\HH(x)$ via the algorithm in
Appendix \ref{AppC} the sign property suggested in \cite{FGG010} that
$\HH(x)$ might have all coefficients, but the first, negative does not
follow; see the final comments in Appendix \ref{AppC}.

In the case of the geodesic motion near the equatorial circle of the
ellipsoid let $a,b$ be the equatorial and polar radii of an ellipsoid of
revolution. The Hamiltonian for the geodesic motion is

\be H=\frac12 \frac{B^2}{b^2\cos^2\b+ a^2\sin^2\b}+\frac12
\frac{A^2}{a^2\cos^2\b}\Eq{e8.3}\ee
where $A$ is the polar component of the angular momentum and
$\b=\th-\frac\p2$ the polar angle, \cite[Problem 4.11.5]{Ga983}.  Let
$r^2\defi \frac{a^2}{b^2}$, $\e\defi r^2-1$,
$X=\frac{B}{b\sqrt{1+\e\sin^2\b}}, Y=-\frac{A}a\tan\b$, $X_\pm=X\pm i Y$;
then $H=\frac12(X^2+Y^2)+\frac{A^2}{2a^2}$ and:

\be \eqalign{
dB&\wedge d\b=\frac{ab}{A}\frac{\sqrt{1+\e 
\sin^2\b}}{1+\tan^2\b}dX\wedge dY
\cr
&=\frac{ab}{A}\frac{(1+r^2 \tan^2\b)^{\frac12}}{(1+\tan^2\b)^{\frac32}}
dX\wedge dY=\frac{ab}{A}\frac{(1+r^2 (\frac{Ya}{A})^2)^{\frac12}}
{(1+(\frac{Ya}{A})^2)^{\frac32}}
 dX\wedge dY\cr
&
=\frac{ab}{A}
\sum_{h,k}{\frac12\choose h}{-\frac32\choose k} r^{2h} 
(\frac {(X_+-X_-)\,a}{2iA})^{2(h+k)}dX\wedge
  dY\cr
&=\frac{ab}{A}\sum_{n=0}^\infty \Big(\frac{X_+X_- a^2}{4A^2}\Big)^{n}
{2n\choose n}
\sum_{h+k=n} {\frac12\choose h}{-\frac32\choose k} r^{2h}dX\wedge
  dY+{\bf 0} \cr}
                 \Eq{e8.4}\ee
Therefore the normal form is $H=\frac{A^2}{2a^2}+\frac{2A^2}{a^2}\HH(x)$
with $x=\frac{a}{4A b}(p^2+q^2)$ and $\HH(x)$ the inverse function to $x=z
  C^2(z)$ with $C^2(z)\defi\sum_{n=0}^\infty P_n(r^2) z^n$ and

\be P_n(r^2)\defi \frac1{n+1}{2n\choose n}
\sum_{h+k=n} {\frac12\choose h}{-\frac32\choose k} r^{2h}\Eq{e8.5}\ee
which is interesting as it shows that the location of the zeros on the unit
circle is a peculiarity of the particular case of the rigid body. 

{\it Nevertheless} a numerical investigation for $n\le 1000$ with an
algebraic manipulator indicates, as a reasonable conjecture, that the zeros
become asymptotically located on the unit circle: for $n$ large they are
within an annulus of radii $r'^2_n<r''^2_n$ with:

\be \cases{ {r'}^2\ge_{n\to\infty} 1+\frac{a'}{n^{c'}}\cr
r''^2\le_{n\to\infty} 1+\frac{a''}{n^{c''}}\cr},\qquad 
c'\simeq 1, c''\simeq .88\Eq{e8.6}\ee
with $\log a''\simeq 1.992, c''\simeq0.880$, $\log
a'\simeq0.462, c'\simeq1.000$. 
%The identities $\sum_{h=0}^n
%{\frac12\choose h}{-\frac32\choose n-h}=(-1)^n$, and $\sum_{h=0}^n
%h{\frac12\choose h}{-\frac32\choose n-h}=\frac{(-1)^{n+1}n}{2}$ hold.

\appendix

\def\inizA{\setcounter{equation}{0}
\renewcommand{\theequation}{\Alph{section}.\arabic{equation}}
}

\inizA

%%%%%%%%%%%%%%%%%%%%%%%%%%%%%%%%%%%%%%%%%%%%%%%%%%%%
%%%%%%%%%%%%%%%%%%%%%%%%%%%%%%%%%%%%%%%%%%%%%%%%%%%%
%%%%%%%%%%%%%%%%%%%%%%%%%%%%%%%%%%%%%%%%%%%%%%%%%%%%
%%%%%%%%%%%%%%%%%%%%%%%%%%%%%%%%%%%%%%%%%%%%%%%%%%%%
\section{Hyperbolic and elliptic coordinates: details}
\label{AppA}
%%%%%%%%%%%%%%%%%%%%%%%%%%%%%%%%%%%%%%%%%%%%%%%%%%%%
%%%%%%%%%%%%%%%%%%%%%%%%%%%%%%%%%%%%%%%%%%%%%%%%%%%%
%%%%%%%%%%%%%%%%%%%%%%%%%%%%%%%%%%%%%%%%%%%%%%%%%%%%
%%%%%%%%%%%%%%%%%%%%%%%%%%%%%%%%%%%%%%%%%%%%%%%%%%%%

In this appendix we give details for the derivation of the formulae in Sec.
\ref{sec3}. 

Let $\sqrt{-1}=-i$ so that $g_a=\sqrt{a} |g_a|$; for $a=+$ Eq.\equ{e2.3}
describe the separatrix branches emerging from the proper rotation around
the axis $\V i_1$ and for $a=-$ they describe the motion near the stable
rotation around the axis $\V i_2$. For $U-A^2/2J_a>0$ close to $0$ and from
the definitions of the Jacobi elliptic integrals, see \cite[2.616]{GR965},
it follows (choosing the sign $+$ in the Eq.\equ{e2.5}, for instance):
\be% \eqalign{&
\frac{\sqrt{1+as^2_a}\,\,\sin\b(t)}{\sqrt{1+a s^2_a\,\sin^2\b(t)}}
={\rm sn}(g_at,k_a)%\cr& 
\otto \
\sin\b(t)=\frac{{\rm sn}(g_at,k_a)
}{\sqrt{1+a s^2_a{\rm cn}^2(g_at,k_a)}}.%\cr}
\Eq{eA.1}\ee
and define $h_a,h'_a$ in terms of $k_a,k'_a$ so that $k_+,k'_+,h_-,h'_-\in
(0,1)$ as:

\be h'^2_a=\frac1{k_a^2},\qquad h^2_a=
1-\frac1{k_a^2}
\Eq{eA.2}\ee
Then (\cite[8.153.3]{GR965} for $a=+$ and \cite[8.153.1-8]{GR965} for $a=-$
after correcting the obvious typo in \cite[8.153.8]{GR965} and by $k'_-=-i
\sqrt{k_-^2-1}$):
\bea
&&
\kern-4mm B(t)=\frac{{b_a}}{{\rm dn}(u_a,k_a)}= b_a\frac{{\rm
      cn}(-iu_a,k'_a)}{{\rm dn}(-iu_a,k'_a)}=
b_a {\rm cn}(-iu_a h'^{-1}_a,h_a),\nn\\
&&\kern-4mm \sin^2\b(t)=\frac{{\rm
      sn}^2(u_a,k_a)}{{1+a s^2_a {\rm cn}^2(u_a,k_a)}}
=\frac{-\frac{{\rm sn}^2(-iu_a,k'_a)}{{\rm cn}^2(-iu_a,k'_a)}}
{1+as_a^2\frac1{{\rm cn}^2(-iu_a,k'_a)}}
                                  \Eq{eA.3}\\
&&\kern-4mm \dot\b(t)=\frac{(1+a s_a^2)\lis g_a{\rm dn}(u_a,k_a)}{1+a
  s_a^2{\rm cn}^2(u_a,k_a)}=\lis g_a
\frac{{\rm dn}(-iu_a,k'_a) {\rm
    cn}(-iu_a,k'_a)}{1-\frac1{1+a s^2_a}{\rm sn}^2(-iu_a,k'_a) }
\nn\eea
where the last relation is derived from the equations of motion and 
the previous expression for $\sin^2\b$. Also the second and third relations
in Eq.\equ{eA.3} could be given expressions involving the moduli
$h_a,h'_a$ so that one could choose to use the formulae with $k_a$ for $a=+$
and with $h_a$ for $a=-$.

Remark also that, \cite[8.128.1]{GR965},\cite[8.128.3]{GR965},
$\KJ(k')=h'\KJ(h)$, and also that $\frac{h'_-|k'_-|}{h_-}=1$.

Notice that $\b(0)=0$ and if $ (2U -A^2/J_a) \to 0^+$ it is $k^2_\pm\to1^\mp$,
$a\,g^2_a\to \wt J_a^{-1} J_{12}^{-1}A^2$, $b_a\to 0^+$ and motions with this
energy are ``like'' the motions close to the separatrix or,
respectively, close to equilibrium of a pendulum.

For $(2U-A^2/J_a)\,\le0$, or for $k_a>1$ hence $k'_a$ imaginary, all the
above relations have to be interpreted via suitable analytic continuations.
Some of the following formulae become singular as $U\to A^2/2J_a$, but the
singularity is only apparent and it will disappear from all relevant
formulae derived or used in the following.

\*
Denoting ${\rm cn}={\rm cn}(u_a,k_a)$,
${\rm cn}'={\rm cn}(-i u_a,k'_a)$, ${\rm dn}'={\rm dn}(-i u_a,k'_a)$,
${\rm sn}'\defi$ ${\rm sn}(-i u_a,k'_a)$, from the equation of motion (and
${\rm dn}^2\cdot$ $\equiv 1-k^2_a {\rm sn}^2\cdot$) it
follows
\be\eqalign{
\dot\b=&\frac{\lis g_a}{{\rm sn}(-iv,k'_a)}
\frac{{\rm sn}(-iv,k'_a) {\rm
        dn}(-iu_a,k'_a){\rm cn}(-iu,k'_a)}{1-\frac 1{1+a s_a^2}{\rm
    sn}^2(-i u_a,k'_a)}\cr}
                             \Eq{eA.4}
\ee
identically for all $v$: convenient choice of $v$ is $v=v_a$ with
$k'^2_a{\rm sn}^2(-iv,k'_a)=(1+a s_a^2)^{-1}$, {\it i.e.} ${\rm
  sn}(-iv_a,k'_a)=\frac1{\sqrt{1+ar^2}}$ and $-i v_a\defi {\rm am}({\rm
  arcsin}\frac1{\sqrt{1+a r_a^2}},k'_a)$, then (notice that
$g_a^2k'^2_a=\lis g_a (1+a r_a^2)$):
\be \eqalign{
\dot\b&=\frac12 \lis g_{a} (1+a r^2)^{\frac12}
\sum_{\h=\pm1}\h \ {\rm sn}(-i(u_a+\h v_a),k_a')
\cr
}                                                 \Eq{eA.5}\ee
which follows by the addition
formulae, \cite[8.156.1]{GR965}, from the last of Eq.\equ{e2.5}.

If $\a_a\defi{\rm arcsin}\frac1{\sqrt{1+ar_a^2}}$, {\it i.e.}
$e^{i\a_a}=\frac{i+\sqrt{a r_a^2}}{\sqrt{1+ar_a^2}}$, it is $-i
v_a=\int_0^{\a_a}\frac{d\th}{\sqrt{1-k'^2_a\sin^2\th}}$ and the integral
can be evaluated by series, \cite[2.511.2]{GR965} and
\cite[8.113.1]{GR965}, leading to Eq.\equ{e3.4},\equ{e3.5}.
As remarked in Sec.\ref{sec3} $|\g_+|=1,\ \g_-=i e^\l$ with $\l$ real and
$\lis\g_-$ real, with $\l$ real.

It is useful to check that if $I=\sqrt{4 \wt J_+\wt J_-}$:
\be\eqalign{
& g_{0,a}= \frac{\p|g_a|}{2\KJ(k'_a)},\quad
\frac{2\p b_a}{k'_a \KJ(k'_a)}=2 I \frac{g_{0,a}}{\sqrt{a}},
%\cr&
\quad
 \frac{\p\lis g_a(1+r_a^2)^{\frac12}}{2 |k'_a|
\KJ(k'_a)}=g_{0,a},\cr}\Eq{eA.6}
\ee
From Eq.\equ{e2.6}, via
\cite[8.146.1]{GR965}  and \cite[8.146.2]{GR965}  and
$g_{0,\pm}$ in Eq.\equ{e2.4}, we get:
\bea
&& \dot \b_a= \frac{g_{0,a}}{\sqrt a}
\sum_{k=0}^\infty\sum_{n=1}^\infty
 \sum_{\s,\h=\pm1} \Big(\frac{\h\s}{i}\Big)
\Big(\sqrt{x'_a} (ax'_a)^k\frac{\sqrt a}{{\sqrt a}^{\h\s}}
e^{\s\sqrt{a}\, g_{0,a}t}
\lis\g_{a}^{\h\s}\Big)^{2n-1}
\nn\\
&&\kern5mm= \frac{d}{dt}\frac1a
\sum_{k=1}^\infty \sum_{\s,\h=\pm1}\frac{\h^{\frac{1+a}2}
  \s} {i} {\rm arctanh}(p_\s (ax'_a)^k \lis\g_a^\h)     \qquad{\rm if}
                \Eq{eA.7}\\
&&p'_\s\defi \sqrt{x'_a} e^{\s{\sqrt{a}} g_{0,a}t},\qquad x'_a\equiv
p'_+p'_-
\nn\eea
Therefore this implies:
\be\b(t)_a=a\sum_{k=0}^\infty\sum_{\s,\h=\pm1}
\frac{
\h^{\frac{1+a}2} \s}i\, a^k\,
\,{\rm arctanh}\,
(x'^k_a p'_\s\lis\g_a^{\h})
                        \Eq{eA.8}\ee
{\it i.e.} the second of Eq.\equ{e3.7}. Likewise from the first of
Eq.\equ{e2.3} we derive:
\be\eqalign{
B_a=&I \frac{g_{0,a}}{\sqrt a}
\sum_{\m=\pm1}\sum_{n=1,k=0}^\infty (-1)^{k-1}
\Big((\sqrt{a x'_a})\,(a x'_a)^{k} e^{\s g_{0,a}\sqrt{a} t}\Big)^{2n-1}
\cr
=&-a I g_{0,a}\sum_{\s=\pm}\sum_{m=0}^\infty
\frac{p'_\s (ax'_a)^m}{1+a (p'_\s x'^m_a)^2}, 
\cr           }   \Eq{eA.9}\ee
via \cite[8.146.11]{GR965}, hence we obtain the first of Eq.\equ{e3.7}.
\*

\0{\it Remark:} via identities relating Jacobian elliptic functions it
would be possible to avoid considering functions that are defined by
analytic continuations, for instance when the modulus $k^2_a>1$ or
$k'^2_a<0$, and reduce instead to considering only cases with the
``simple'' arguments (expressing functions by elliptic functions of the
arguments $h,h'$, see for instance the first of Eq.\equ{eA.3}, when
necessary): we avoid this as it would hide the nice property that the
stable and unstable cases are in a sense analytic continuation of each
other.  \*

Summarizing the transformation of coordinates $(B,\b)\otto (p'_+,p'_-)$ in
Eq.\equ{e3.7} is such that $t\to p'_\s e^{\s \sqrt{a}\,g_{0,a}t}$
generates a solution of the equations of motion (if $p'_\pm>0$).

%%%%%%%%%%%%%%%%%%%%%%%%%%%%%%%%%%%%%%%%%%%%%%%%%%%%%%%%%%
%%%%%%%%%%%%%%%%%%%%%%%%%%%%%%%%%%%%%%%%%%%%%%%%%%%%%%%%%%

\def\SEC{\small Roots}
\section{Polynomials roots}\label{AppB}
\inizA
%%%%%%%%%%%%%%%%%%%%%%%%%%%%%%%%%%%%%%%%%%%%%%%%%%%%%%%%%%
%%%%%%%%%%%%%%%%%%%%%%%%%%%%%%%%%%%%%%%%%%%%%%%%%%%%%%%%%%

The proof of the theorem is in the literature: first remark (a classical
result, see \cite[p.107,301]{PS978}) that if $a_0\ge a_1\ge\ldots \ge
a_m\ge0$, and $Q_m(z)=\sum_{k=0}^m a_k z^k\not\equiv0$, then $Q_m(z)$ can
only vanish if $|z|\ge1$. Because
\be\eqalign{
|(1-z)&Q_m(z)|=|a_0+(a_1-a_0)z+\ldots+(a_{n}-a_{n-1})z^n+a_n z^{n+1}|\cr
&\ge a_0-(a_0-a_1)|z|-\ldots-(a_{n-1}-a_n) z^n-a_n |z|^{n+1}\cr
&> a_0-(a_0-a_1)-\ldots-(a_{n-1}-a_n)-a_n=0\cr}\Eq{eB.1}\ee
Following word by word the proof in \cite{Ch995}, 
write the  polynomials $P_n(z),\,n\ge1$
\bea
&&a_0+a_1 z+\ldots a_n z^n+a_n z^{n+1}+\ldots+a_0 z^{2n+1},\qquad{\rm or}
\Eq{eB.2}\\
&&a_0+a_1 z+\ldots a_{n-1} z^{n-1}+\frac12 a_n z^n +\frac12 a_n z^n
+a_{n-1} z^{n+1}+\ldots+a_0 z^{2n}\nn\eea
as $Q_n(z)+z^{n^*} Q_n^*(z)$ with $n^*=n+1$ or $n^*=n$ respectively and 
\be Q_n(z)=\sum_{k=0}^n \a_k
z^k, \qquad Q_n^*(z)\defi z^n \lis{Q_n(\lis z^{-1})}\equiv z^n Q_n(z^{-1})
,\Eq{eB.3}\ee 
with $\a_0\ge \a_1\ge\ldots\ge\a_n$ (with $\a_k=a_k$ for $k<n$
and $\a_n=a_n$ or $\a_n=\frac12 a_n$).

Hence $Q_n(z)=a_0\prod (z-\frac1{z_k})$ with $|z_k|\le1$ by the above
remark, and $Q^*_n(z)=a_0 z^n\prod (\frac1z-\frac1{\lis z_k})$.

If $\z$ is a zero for $P_n$ it is $Q_n(\z)=-\z^{n^*}Q^*_n(\z)$ or
\be \frac{ |\z|^{n^*}\prod |1-{\z\lis z^{-1}_k}|}{\prod |\z-{
    z^{-1}_k}|}=\frac{ |\z|^{n^*}\prod |\z-\lis z_k|}
{\prod |1-\z{z_k}|}%\prod|\frac{z_k}{\lis z_k}|
=1,\Eq{eB.4}\ee
possible only if $|\z|=1$: because if $|a|<1$ then it is
$|\frac{z-a}{1-\lis a z}|>1$ for $|z|>1$ and $<1$ for $|z|<1$ (and
$n^*>0$); or if $|z_k|\equiv1$ Eq.\equ{eB.4} means $|\z|^{n^*}=1$.

%%%%%%%%%%%%%%%%%%%%%%%%%%%%%%%%%%%%%%%%%%%%%%%%%%%%%%%%%%
%%%%%%%%%%%%%%%%%%%%%%%%%%%%%%%%%%%%%%%%%%%%%%%%%%%%%%%%%%

\def\SEC{\small Tree expansion}
\section{Tree expansion}\label{AppC}
\inizA
%%%%%%%%%%%%%%%%%%%%%%%%%%%%%%%%%%%%%%%%%%%%%%%%%%%%%%%%%%
%%%%%%%%%%%%%%%%%%%%%%%%%%%%%%%%%%%%%%%%%%%%%%%%%%%%%%%%%%

To invert the relation $x=z\,\wh C(z)^2$ let $f(z)\defi z\,\wh
C(z)^2-z=\frac{1+r^2}2 z^2+\ldots\equiv \sum_{n=2}^\infty f_n\,z^n$, see
Eq.\equ{e6.4}, with $f_n=P_{n-1}$ a polynomial in $r^2$ of degree
$n-1$. Consider the equation
\be z=x-f(z),\ {\rm with}\ z=x+h(x),\quad{\rm or}\quad h(x)=-
f(x+h(x)) \Eq{eC.1}\ee
and look for a solution $h(x)=\sum_{k=2}^\infty h_k x^k$.

Let $\th$ be a tree graph with a root and nodes $v$ into each of which
merge $s_v=0,1,2,3,4,\ldots$ branches oriented towards the root; each node
$v$ carries a label $n_v=2,\ldots$ with the restriction $\sum_v
s_v=k$ unless $v$ is an end node in which case $n_v=1$ and $s_v=0$. 

Such labeled tree graphs will form a family $\Th(k)$ of trees identified by
the rule that two trees are regarded as identical if they can be overlapped
by pivoting the branches avoiding that any two branches overlap in the
course of the process.

Define the ``value'', ${\rm Val}(\th)$, of $\th$ as $ {\rm
  Val}(\th)=\s(\th)\prod_{v\in\th} {n_v\choose s_v}f_{n_v}$ with
$\s(\th)\defi\prod_{v,n_n>1}(-1)$ and $f_1=1$. Notice that if he
coefficients defining $f_n$, $n\ge2$, are all {\it negative} then ${\rm
  Val}(\th)\ge0$ for all $\th$.  Then
\be h(x)=\sum_{k=1}^\infty \sum_{\th\in\Th(k)}{\rm
  Val}(\th) x^k\Eq{eC.2}\ee
as it is checked by induction (for instance). 

In our case $f_n=P_{n-1}(r^2)$ and the contributions of the value of a tree
$\th$ to the coefficient of $x^k$ is a product of symmetric polynomials
$P_{n_i}(r^2)={2 n_i\choose n_i}\sum_{h=0}^{n_i} {2 h\choose h} {2 (n_i-h)
\choose (n_i-h)} r^{2h}$ with symmetrically decreasing coefficients and
of total degree $p-1$ in $r^2$; consequently the polynomials in $r^2$ being
sums of products of polynomials enjoying the $z\otto z^{-1}$ symmetry
property do enjoy it as well. However positivity and monotony is in general
lost.

Notice that Eq.\equ{e6.7}, \equ{e8.2} suggest that the coefficients of
  $x^n$ with $n>1$ are negative polynomials $p_n(r^2)$: in fact $-p_n(x)$
  are either squares of $(1-r^2)$ times polynomials with positive
  coefficients for $n$ odd or, for $n$ even, $(1-r^2)^2$ times $(1+r^2) \wt
  p_n(r^2)$ which after performing the multiplication become a polynomial
  with positive coefficients.  This property, however, does not follow from
  the above tree expansion and we wonder whether it continues to be true at
  the higher orders.

\* \0{\bf Acknowledgment:} We are indebted to G. Gentile, A. Giuliani for
discussions and to A. Giuliani for the suggestions quoted in Sec.\ref{sec6}.

\small

\bibliographystyle{unsrt}

\end{document}